\documentstyle[12pt, epsfig]{article}
\begin{document}
\vspace*{-1cm}

\normalsize
\hspace*{10cm}\parbox{15cm}{ADP-94-5/T147                       \\
Physics Letters B 340 (1994) 115-121}
\vspace*{2.0cm}

\Large
\begin{center}
Drell-Yan Processes as a Probe of Charge Symmetry               \\
Violation in the Pion and the Nucleon                           \\
\end{center}

\vspace*{0.4cm}

\normalsize
\begin{center}
{\normalsize J.T. Londergan}                                    \\
{\small\it Department of Physics and Nuclear Theory Center      \\
Indiana University                                              \\
Bloomington IN 47408 USA},                                      \\

\vspace*{0.4cm}
{\normalsize G.T. Garvey}                                       \\
{\small\it Los Alamos National Laboratory                       \\
Los Alamos, NM 87544 USA},                                      \\

\vspace*{0.4cm}
{\normalsize G.Q. Liu, E.N. Rodionov and A.W. Thomas}           \\
{\small\it Department of Physics and Mathematical Physics       \\
University of Adelaide                                          \\
Adelaide, S.A., 5005, Australia}
\end{center}

\vspace*{1cm}
\large
\begin{center}
Abstract
\end{center}
\normalsize

\hspace*{-0.5cm}
We extend earlier investigations of charge symmetry violation in the valence
quark distributions of the nucleon, and make similar estimates for the pion.
The sensitivity of pion-induced Drell-Yan measurements to such effects is then
examined. It is shown that combinations of $\pi^+$ and $\pi^-$ data on
deuterium
and hydrogen are sensitive to these violations, and that the pion and nucleon
charge symmetry violating terms separate as a function of $x_\pi$ and $x_N$
respectively. We estimate the background terms which must be evaluated to
extract charge symmetry violation.

\vspace*{3cm}


\vspace*{1cm}
{\small e-mail:\\
\hspace*{2.0cm}\parbox{15cm}{londergan@venus.iucf.indiana.edu\\
garvey@lampf.lanl.gov\\
gliu,$\:$ erodiono,$\:$ athomas@physics.adelaide.edu.au
}}
\newpage

At the present time the flavor structure of the nucleon is a topic of intense
interest [1-8]. This is largely a consequence of unexpected experimental
results, such as the discovery by the New Muon Collaboration (NMC) of a
violation of the Gottfried sum-rule \cite{nmc} and the so-called ``proton
spin crisis'' \cite{emc,slacspin,revspin} of EMC.  There have also been
recent theoretical calculations of the violation of charge symmetry in the
{\bf valence} quark distributions of the nucleon \cite{sather,rtl}.

In most nuclear systems, charge symmetry is obeyed to within about
one percent \cite{mill}, so one would expect small charge symmetry
violation [CSV] in parton distributions. The theoretical calculations
suggest that there is a CSV part of the ``minority'' valence quark
distributions ($d^p$ or $u^n$), with a slightly smaller violation
in the ``majority'' valence distributions ($u^p$ or $d^n$).  Although both
CSV contributions are rather small in absolute magnitude, the {\bf fractional}
charge symmetry violation in the minority valence quark distributions
$r_{min}(x) \equiv 2(d^p(x) - u^n(x))/(d^p(x) + u^n(x))$ can be large,
because at large momentum fraction $x$, $d^p(x)/u^p(x) << 1$. Rodionov
et al.\ \cite{rtl} predicted charge symmetry violation as large as $5-10$\%
for the ratio $r_{min}(x)$, in the region $x > 0.5$. The relative size of
these CSV effects might require a change in the standard notation
for parton distributions \cite{fec} in this region; in addition, Sather
\cite{sather} showed that CSV effects of this magnitude could significantly
alter the value of the Weinberg angle extracted from neutral and charged
current neutrino interactions.

Since (in this particular region of Bjorken $x$) we predict fractional CSV
violations as large as $5-10$\%, it is important to explore experiments which
would be sensitive to the relative minority quark distributions in the neutron
and proton. Observation of a CSV effect at this level would reinforce
confidence in our ability to relate quark models to measured quark-parton
distributions -- and hence to use deep inelastic scattering as a real probe
of the non-perturbative aspects of hadron structure \cite{mt3}. Drell-Yan
processes have proven to be a particularly useful source of information on
the anti-quark distributions in nuclei \cite{garvey}. If one uses beams of
pions, and concentrates on the region where Bjorken $x$ of the target quarks
is reasonably large, then the annihilating quarks will predominantly come
from the nucleon and the antiquarks from the pion.  Furthermore, for
$x \geq 0.4$ to good approximation the nucleon consists of three valence
quarks, and the pion is a quark-antiquark valence pair -- in particular,
$\pi^+$ contains a valence $\bar{d}$ and $\pi^-$ a valence $\bar{u}$.
Comparison of Drell-Yan processes induced by $\pi^+$ and $\pi^-$ in this
kinematic region will provide a good method of separately measuring d and u
quark distributions in the nucleon.

Consider the Drell-Yan process in which a quark with momentum fraction
$x_1$ in a deuteron annihilates with an anti-quark of momentum fraction
$x_2$ in a $\pi^+$. Provided that $x_1, x_2 \geq 0.4$, to minimize the
contribution from sea quarks, this will be the dominant process.  Neglecting
sea quark effects, the Drell-Yan cross section will be proportional to:
\begin{equation}
\sigma_{\pi^+D}^{DY} \sim {1 \over 9}\left( d^p(x_1) + d^n(x_1) \right)
 \overline{d}^{\pi^+}(x_2).
\label{eq:pipd}
\end{equation}
The corresponding cross section for $\pi^-$D is :
\begin{equation}
\sigma_{\pi^-D}^{DY} \sim {4\over 9} \left( u^p(x_1) + u^n(x_1) \right)
 \overline{u}^{\pi^-}(x_2),
\label{eq:pimd}
\end{equation}
so that if we construct the ratio, $R^{DY}_{\pi D}$:
\begin{equation}
R^{DY}_{\pi D}(x_1, x_2) =
\frac{4 \sigma_{\pi^+D}^{DY} - \sigma_{\pi^-D}^{DY}}
{\left(4 \sigma_{\pi^+D}^{DY} + \sigma_{\pi^-D}^{DY}\right) /2},
\label{eq:R}
\end{equation}
only charge symmetry violating (CSV) terms contribute. In fact, defining
\begin{equation}
\delta d = d^p - u^n, \quad
\delta u = u^p - d^n, \quad
\delta \overline{d}^\pi = \overline{d}^{\pi^+} - \overline{u}^{\pi^-},
\label{eq:deltas}
\end{equation}
(and recalling that charge conjugation implies
$\overline{d}^{\pi^+}=d^{\pi^-}$
etc.\ ) we
find that, to first order in the small CSV terms, the Drell-Yan ratio becomes:
\begin{eqnarray}
R^{DY}_{\pi D}(x_1, x_2) & = & \left( \frac {\delta d - \delta u}{u^p + d^p}
 \right)  (x_1) + \left( \frac{\delta \overline{d}^\pi}{\overline{d}^{\pi^+}}
 \right)(x_2), \nonumber \\
           & = & R^N_{\pi D}(x_1) + R^{\pi}(x_2),
\label{eq:Rfinal}
\end{eqnarray}

Equation (\ref{eq:Rfinal}) is quite remarkable in that only CSV quantities
enter, and there is a separation of the effects associated with the nucleon
and the pion. Because $R^{DY}_{\pi D}$ is a ratio of cross sections one
expects a number of systematic errors to disappear -- although the fact that
different beams ($\pi^+$ and $\pi^-$) are involved means that not all
such errors will cancel. This certainly needs further investigation, since
$R^{DY}$ is obtained by almost complete cancellation between terms in
the numerator.  The largest ``background'' term, contributions from nucleon
or pion sea, will be estimated later in this letter. However we note that
Equ.\ (\ref{eq:Rfinal}) is not sensitive to differences between the parton
distributions in the free nucleon and those in the deuteron
\cite{bodek,fs,bicker,mst}.  For example, if the parton distributions
in the deuteron are related to those in the neutron and proton by
\begin{equation}
q_i^D(x) = \left( 1 + \epsilon (x) \right) \left( q_i^p(x) + q_i^n(x)
  \right) \quad ,
\label{eq:epsilon}
\end{equation}
then by inspection equ.\ (\ref{eq:Rfinal}) will be unchanged.  Any correction
to the deuteron structure functions which affects the proton and neutron terms
identically will cancel in $R^{DY}$.

It should not be necessary to know absolute fluxes of charged pions to
obtain an accurate value for $R^{DY}$.  The yield of $J/\psi$'s from
$\pi^+-D$ and $\pi^--D$ can be used to normalize the relative fluxes,
since the $J/\psi$'s are predominantly produced via gluon fusion processes.
The gluon structure functions of the $\pi^+$ and $\pi^-$ are identical,
so the relative yield should be unity to within 1\%.

Next we turn to the predictions for the charge symmetry violating terms,
$\delta d, \delta u$ and $\delta \bar{d}^\pi$ which appear in
equ.(\ref{eq:Rfinal}). For the former two there has been an extensive
discussion by Sather \cite{sather} and Rodionov et al.\cite{rtl} from which
there is at least a theoretical consensus that the magnitude of $\delta d$
is somewhat larger than $\delta u$. This is easy to understand because the
dominant source of CSV is the mass difference of the residual di-quark pair
when one quark is hit in the deep-inelastic process. For the minority
quark distribution the residual di-quark is $uu$ in the proton, and $dd$ in
the neutron. Thus, in the difference, $d^p - u^n$, the up-down mass difference
enters twice.  Conversely, for the majority quark distributions the residual
di-quark is a $ud$-pair in both proton and neutron, so there is no contribution
to CSV.

In Fig.\ 1(a) we show the predicted CSV terms for the majority and minority
quark distributions in the nucleon, as a function of $x$, calculated
for the simple MIT bag model \cite{rtl,MIT,adel}. There are, of course,
more sophisticated quark models available but the similarity of the
results obtained by Naar and Birse \cite{birse} using the color dielectric
model suggests that similar results would be obtained in any relativistic
model based on confined current quarks.  The bag model parameters are listed
in this Figure.  The mean di-quark masses are chosen as 600 MeV for the $S=0$
case, and 800 MeV for the $S=1$ case (note that for the minority quark
distributions the di-quark is always in an $S=1$ state).  The di-quark mass
difference, $m_{dd} - m_{uu}$, is taken to be 4 MeV, a rather well determined
difference in the bag model. We note that $\delta u$ is opposite in sign  to
$\delta d$ and, therefore, these two terms add constructively in the Drell-Yan
ratio $R^{DY}$ of equ.\ (\ref{eq:Rfinal}).\\
\begin{picture}(0,0)(0,220)
\mbox{\epsfig{file=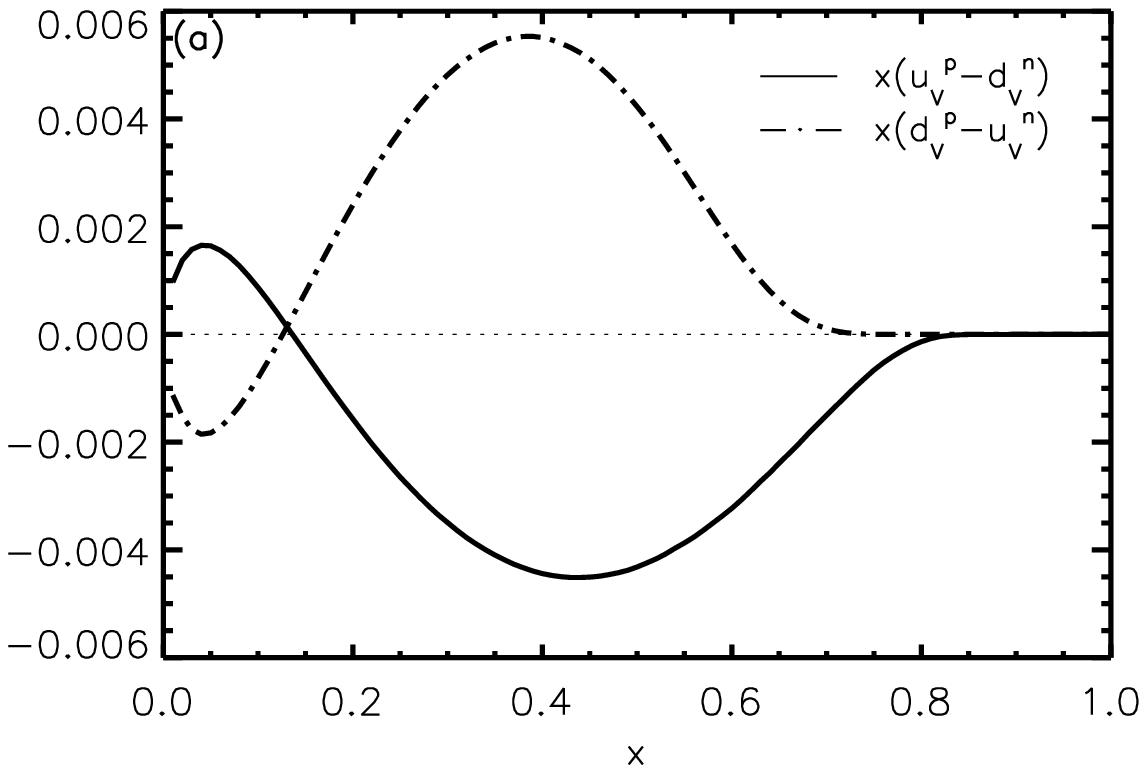,height=12cm,width=18cm,angle=0}}
\end{picture}
\\
\begin{picture}(0,0)(-55,420)
\mbox{\epsfig{file=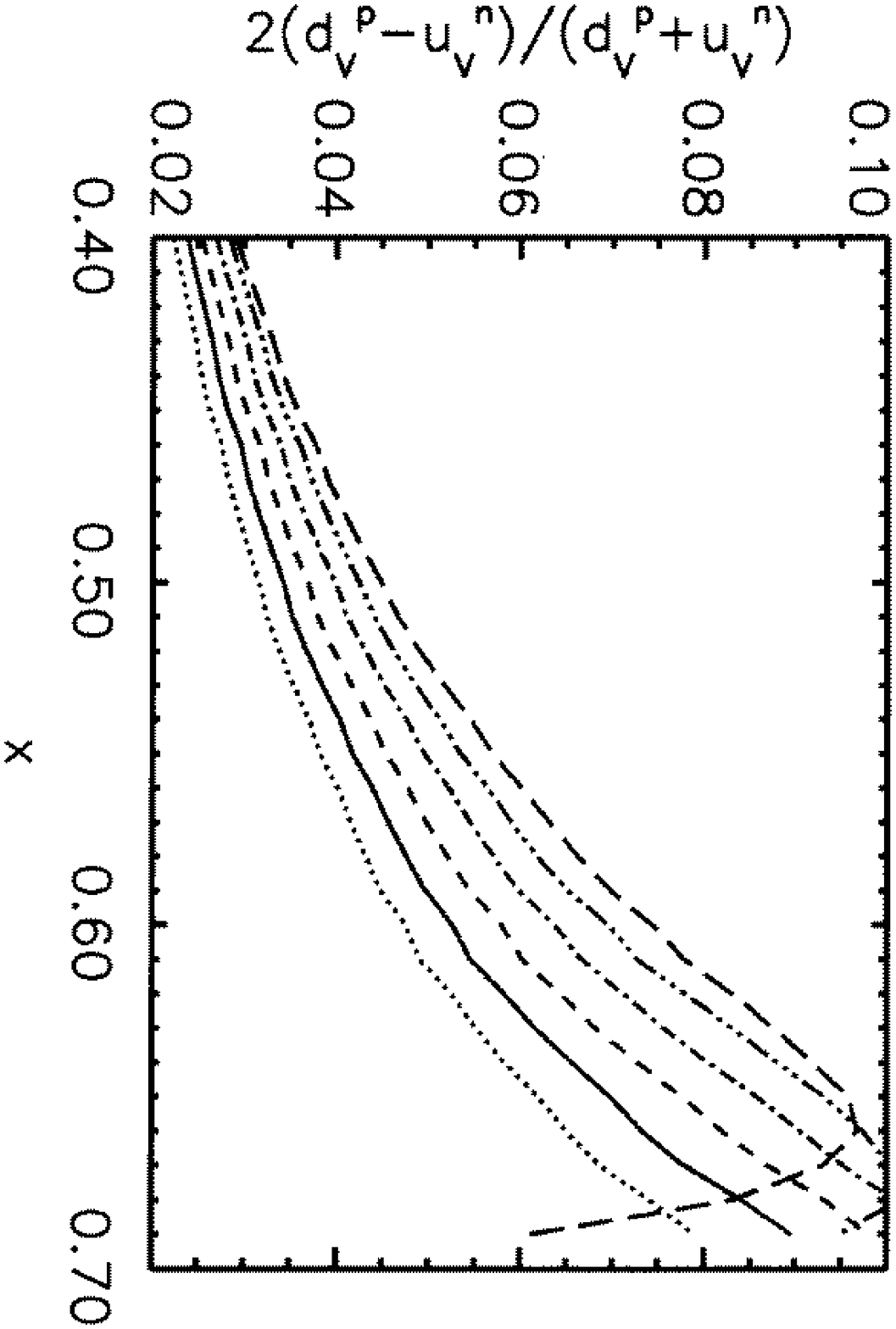,height=7cm,width=13cm,angle=90}}
\end{picture}

\vspace*{15cm}
Fig. 1.\\ \hspace*{2cm}\parbox{13cm}{
(a) Predicted charge symmetry violation (CSV), calculated using the MIT bag
model.  Dashed curve: ``minority'' quark CSV term,
$x\delta d(x) = x\left( d^p(x) - u^n(x) \right)$; solid curve:
``majority'' quark CSV term, $x\delta u(x) = x\left( u^p(x) - d^n(x)\right)$.
(b) Fractional minority quark CSV term, $\delta d(x)/ d^p(x)$, vs.\
$x$, as a function of the average di-quark mass $\overline{m}_d = (m_{uu} +
m_{dd})/2$.  The di-quark mass difference is fixed at $\delta m_d = m_{dd} -
m_{uu} = 4$ MeV.  From top to bottom, the curves correspond to average
di-quark mass $\overline{m}_d =$ 850, 830, 810, 790, 770, and 750 MeV.
The curves have been evolved to $Q^2 = 10$ GeV$^2$.  This quantity is the
nucleonic CSV term for the Drell-Yan ratio $R^{DY}_{\pi N}$ of
Equ.\ (\ref{eq:Rnfinal}).
}

\newpage
In Fig.\ 1(b) we show the fractional change in the minority quark
CSV term, $2(d^p - u^n)/(d^p + u^n)$ vs.\ $x$ for several values of
the intermediate mean di-quark mass.  Although the precise value of
the minority quark CSV changes with mean di-quark mass, the size is
always roughly the same and the sign is unchanged.  This shows that
``smearing'' the mean di-quark mass will not dramatically diminish the
magnitude of the minority quark CSV term (the mean di-quark mass must be
roughly 800 MeV in the $S=1$ state to give the correct $N-\Delta$
mass splitting).

In Fig.\ 2 we show the nucleon CSV contribution, $R^N_{\pi D}(x_1)$,
calculated using the same bag model.  As is customary in these bag model
calculations, this term is calculated at the bag model scale (0.5 GeV for
$R = 0.8$ fm) and then evolved to higher $Q^2$ using the QCD evolution
equations \cite{altpar}.  As the main uncertainty in our calculation is the
mean di-quark mass (the splitting between $S=0$ and $S=1$ is kept at
$200$ MeV \cite{close}), the results are shown for several values of this
parameter. In the region $0.4 \leq x \leq 0.7$, we predict $R^N$ will be
always positive, with a maximum value of about 0.015.  For $x > 0.7$ the
struck quark has a momentum greater than 1 GeV which is very unlikely in a
mean-field model like the bag. As a consequence the calculated valence
distributions for the bag model tends to be significantly smaller than the
measured distributions in this region. In these circumstances one cannot
regard the large, relative charge symmetry violation found in this region as
being reliable and we prefer not to show it. It would be of particular
interest to add $q-q$ correlations which are known to play an important
role as $x \rightarrow 1$ \cite{far}.\\
\begin{picture}(0,0)(0,210)
\mbox{\epsfig{file=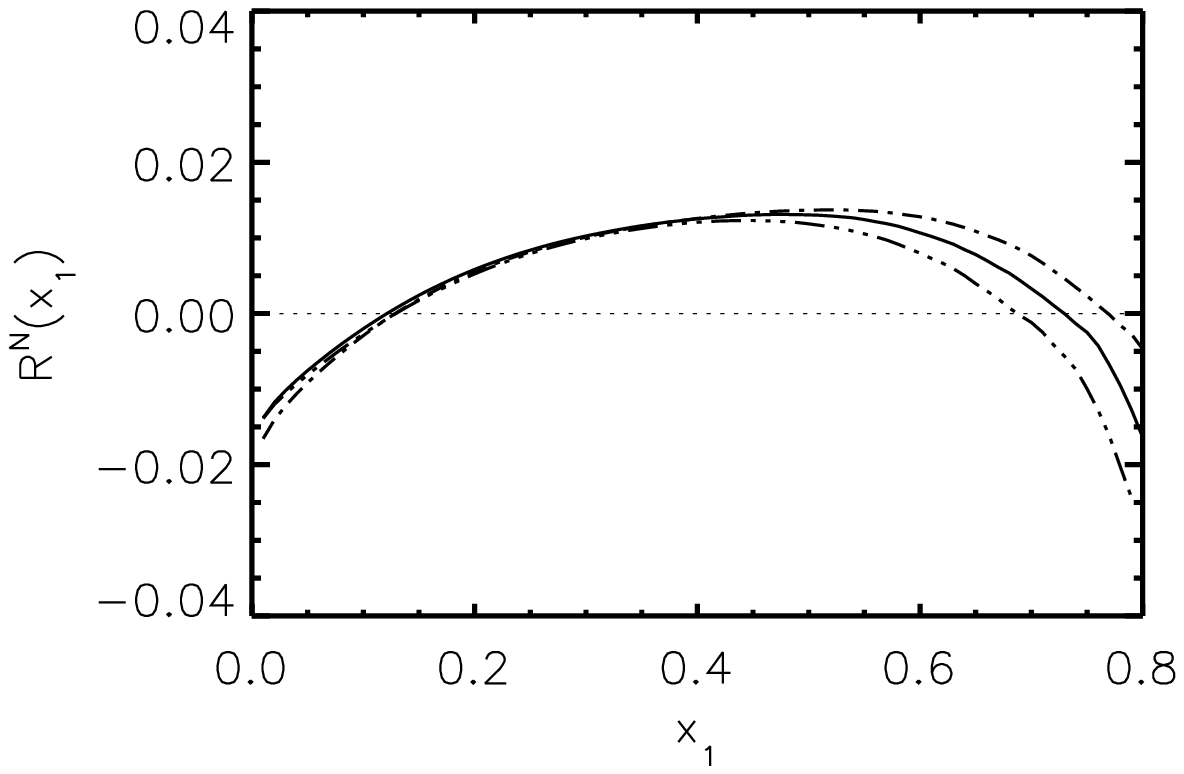,height=12cm,width=18cm,angle=0}}
\end{picture}

\vspace*{6.5cm}
Fig. 2.\\ \hspace*{2cm}\parbox{13cm}{
Contributions from various sources to the nucleonic charge symmetry breaking
term $R^{DY}_{\pi D}$ of equ.\ (\ref{eq:Rfinal}), evolved to $Q^2 = 10$
GeV$^2$,
for pion-induced Drell-Yan ratios on deuterons.  Unless otherwise specified,
parameters are:
$m_d = m_u$ = 0 MeV;   $M^{(n)} = M^{(p)}$ = 938.27 MeV;
$R^{(n)} = R^{(p)}$ = 0.8 fm; $\delta m_d = m_{dd} - m_{uu} = 4$ MeV.
Curves represent different values of average di-quark mass.  Dash-dot:
$\overline{m}_d =$ 750 MeV; solid: 800 MeV; dash-triple-dot: 850 MeV.
}

\vspace*{0.5cm}
For the pion, calculations based on the MIT bag model are really not
appropriate. In particular, the light pion mass means that center of mass
corrections are very large, and of course bag model calculations do not
recognize the pion's Goldstone nature. On the other hand the model of Nambu
and Jona-Lasinio (NJL) \cite{njl} is ideally suited to treating the structure
of the pion, and there has been recent work, notably by Toki and collaborators,
in calculating the structure function of the pion (and other mesons) in this
model \cite{toki1,toki2}. The essential element of their calculation was the
evaluation of the so-called handbag diagram for which the forward Compton
amplitude is:
\begin{equation}
T_{\mu \nu} = i \int \frac {d^4 k}{(2\pi)^4} Tr [\gamma_\mu Q
\frac{1}{\not{k}} \gamma_\nu Q T_{-}] + (T_{+} term),
\label{eq:compton}
\end{equation}
where
\begin{equation}
T_{-} = S_F(k-q, M_1) g_{\pi qq} \tau_{+} i \gamma_5 S_F(k-p-q, M_2)
g_{\pi qq}\tau_{-} i \gamma_5 S_F(k-q, M_1),
\label{eq:T}
\end{equation}
represents the contribution with an anti-quark of mass $M_2$ as spectator to
the absorption of the photon (of momentum $q$) by a quark. In equ.(\ref{eq:T})
$g_{\pi qq}$ is the pion-quark coupling constant, Q the charge of the struck
quark, $p$ the momentum of the target meson, and $S_F$ the quark propagator.
$T_+$ is the corresponding term where the quark is a spectator and the
anti-quark undergoes a hard collision. As in our bag model studies this model
was used to determine the leading twist structure function at some low scale
($0.25$ GeV in this case), and then evolved to high-$Q^2$ using the
Altarelli-Parisi equations \cite{altpar}. The agreement between the existing
data and the calculations for the pion and kaon obtained in ref.(\cite{toki1})
was quite impressive.

\begin{picture}(0,0)(25,270)
\mbox{\epsfig{file=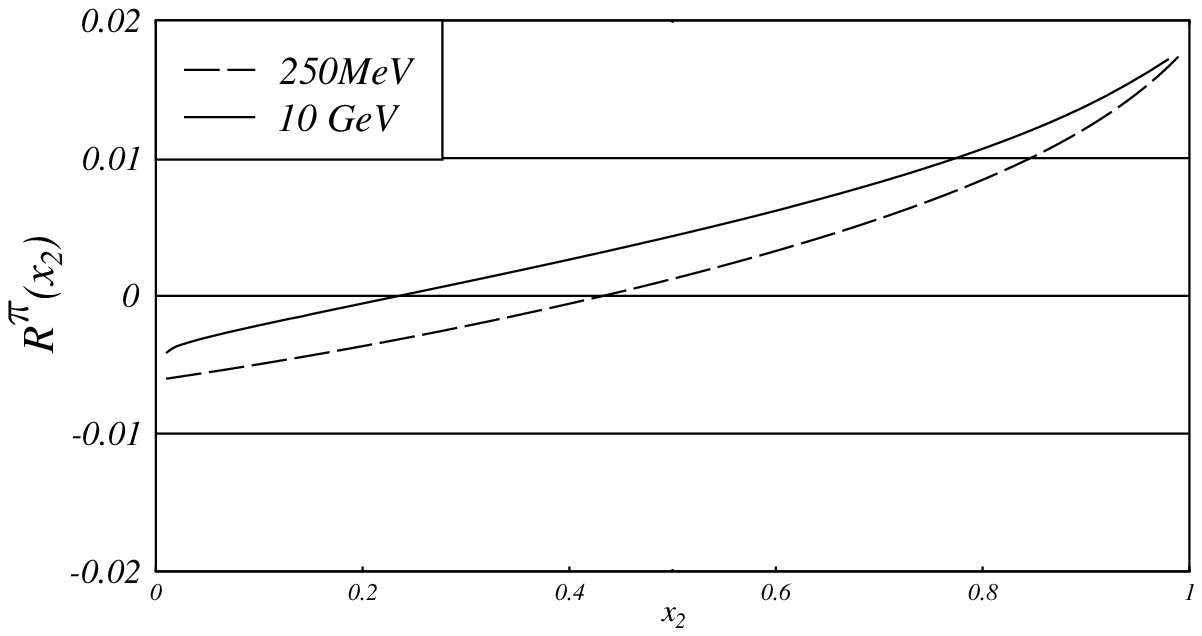,height=12cm,width=18cm,angle=0}}
\end{picture}

\vspace*{7.5cm}
Fig. 3.\\ \hspace*{2cm}\parbox{13cm}{
 Calculation of the pion CSV contribution $R^\pi$ of equ.\
(\ref{eq:Rfinal}), using the model of Refs.\ \cite{toki1,toki2}, for average
nonstrange quark mass 350 MeV, and mass difference $m_d - m_u = 3$ MeV. Dashed
curve: $R^\pi$ evaluated at the bag scale, $\mu = 250$ MeV; solid curve: the
same quantity evolved to $Q^2 = 10$ GeV$^2$.
}

\vspace*{0.5cm}
In the light of the successful application of the NJL model to the structure
functions of the pion and the kaon, where the dominant parameter is the mass
difference between the constituent strange and non-strange quarks (assumed to
be about $180$ MeV), it seems natural to use the same model to describe the
small difference $\delta \bar{d}^\pi$ (c.f. equ.(\ref{eq:deltas})) arising
from the 3 MeV constituent mass difference of the u and d\cite{mill}.  We have
carried out such a calculation. Figure 3 shows the pionic contribution to the
CSV Drell-Yan ratio, $R^{\pi}(x_2)$, corresponding to this mass difference
and an average non-strange quark mass of $350$ MeV -- as used in
ref.(\cite{toki1,toki2}).
The dashed curve is the result at the bag scale, and the solid curve shows
the result evolved to $Q^2 = 10$ GeV$^2$. For $x_2 > 0.3$, we predict that
$R^\pi$ will be positive and increase monotonically, reaching a value of
about 0.01 at $x_2 = 0.8$ and increasing to about 0.02 as $x_2 \rightarrow 1$.
(We note that there is some ambiguity in translating the usual Euclidean
cut-off in the NJL model into the cutoff needed for deep inelastic scattering.
In order to study charge symmetry violation we are particularly concerned to
start with a model that gives a good description of the normal pion structure
function. For this reason we have chosen to follow  the method used in
ref.\cite{toki1} rather than ref.\cite{toki2}.)

Up to this point we have neglected the nucleon and pion sea quark
contributions.
Since the Drell-Yan ratios arising from CSV are very small (viz. Figures 2 and
3), even small contributions from sea quarks could make a substantial effect.
The dominant contribution will arise from interference between one sea quark
and one valence quark.  Assuming charge symmetry for the quark distributions,
and using the same form for the quark and antiquark distributions in the pion,
the sea-valence contribution to the Drell-Yan ratio of Eq.\ (\ref{eq:Rfinal})
has the form
\begin{eqnarray}
R^{SV}_{\pi D}(x_1, x_2) &=&  \frac{3 \sigma_{\pi D}^{SV}(x_1, x_2)}
 {\left({4\over 9}\overline{d}^{\pi}_v(x_2)\left( u^p_v(x_1) + d^p_v(x_1)
 \right) + {5 \over 2}\sigma_{\pi D}^{SV}(x_1, x_2)\right) };
  \nonumber \\
\sigma_{\pi D}^{SV}(x_1, x_2) &\equiv& {5\over 9} \left[ 2\pi_v(x_2)u^p_s(x_1)
 + \pi_s(x_2)\left( u^p_v(x_1) + d^p_v(x_1) \right) \right].
\label{eq:Rsv}
\end{eqnarray}
Unlike the CSV contributions of Eq.\ (\ref{eq:Rfinal}), the sea-valence
term does not separate.  In Figure 4 we show the sea-valence term as a
function of $x_1$ and $x_2$ using recent phenomenological nucleon and pion
parton distributions \cite{approx}.  The sea-valence contribution, although
extremely large at small $x$, decreases rapidly as $x$ increases.  For
$x \geq 0.5$, the sea-valence term is no larger than the CSV ``signal''.
With accurate phenomenological nucleon and pion quark distributions, it
should be possible to calculate this contribution reasonably accurately
(the main uncertainty is the magnitude of the pion sea).
For smaller values ($x_1 \approx x_2 \approx 0.4$, where the background
dominates, we could use the data to normalize the pion sea contribution;
we should then be able to predict the sea-valence term rather accurately for
larger $x$ values, where the CSV contributions become progressively more
important.  We could also exploit the very different dependence on $x_1$ and
$x_2$ of the background and CSV terms. We conclude that the CSV terms could
be extracted even in the presence of a sea-valence ``background''.  We
emphasize that our proposed Drell-Yan measurement would constitute the first
direct observation of charge symmetry violation for these quark distributions.

The Drell-Yan CSV ratio $R^{DY}_{\pi D}$ of Eq.\ (\ref{eq:Rfinal}) is the
sum of the nucleon CSV term of Fig.\ 2 and the pion term of Fig.\ 3,
at the respective values of Bjorken $x$.  Since both quantities are positive,
they will add to give the experimental ratio.  Despite the fact that the
fractional minority quark CSV term is as large as 10\% (c.f.\ Fig.\ 1(b)),
the nucleonic CSV ratio $R^N_{\pi D}$ is more like 1-2\%.  This is because
$\delta d$ in equ.\ (\ref{eq:Rfinal}) is divided by $u^p + d^p$ and since
$d^p(x) << u^p(x)$ at large $x$ the nucleon CSV term is significantly
diminished.  A much larger ratio could be obtained by comparing the
$\pi^+ -p$ and ``$\pi^- -n$'' Drell-Yan processes through the ratio :
\begin{equation}
R^{DY}_{\pi N}(x_1, x_2) =
\frac{4 \sigma_{\pi^+p}^{DY} + \sigma_{\pi^-p}^{DY} - \sigma_{\pi^-D}^{DY}}
 {\left(4 \sigma_{\pi^+p}^{DY} - \sigma_{\pi^-p}^{DY}
  + \sigma_{\pi^-D}^{DY}\right) /2}.
\label{eq:Rpin}
\end{equation}
To first order in the small CSV quantities, this ratio can be written:
\begin{eqnarray}
R^{DY}_{\pi N}(x_1, x_2) & = & \frac {\delta d }{d^p}
 (x_1) + \left( \frac{\delta \overline{d}^\pi}{\overline{d}^{\pi^+}} \right)
(x_2), \nonumber \\
           & = & R^N_{\pi N}(x_1) + R^{\pi}(x_2).
\label{eq:Rnfinal}
\end{eqnarray}
Once again, the ratio separates completely in $x_1$ and $x_2$, and the pion
CSV term is identical with equ.\ (\ref{eq:Rfinal}).  However, the nucleon CSV
term is much larger -- in fact, it is precisely the ratio given in Fig.\ 1(b),
so we expect CSV effects at the 5-10 \% level for this quantity.

\begin{picture}(0,0)(-25,230)
\mbox{\epsfig{file=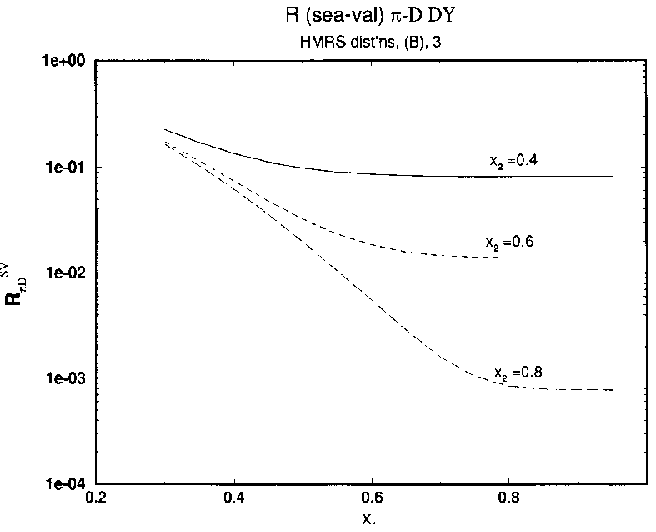,height=8cm,width=12cm,angle=270}}
\end{picture}

\vspace*{8.5cm}
Fig. 4.\\ \hspace*{2cm}\parbox{13cm}{
 The sea-valence contribution $R^{SV}_{\pi D}$ to pion-induced Drell-Yan
ratios on deuterons, as a function of nucleon $x_1$.  Solid curve: pion
$x_2 = 0.4$; dashed curve: $x_2 = 0.6$; long-dashed curve: $x_2 = 0.8$.
This quantity was calculated using nucleon and pion parton distributions of
Ref.\ \cite{approx}.
}

\vspace*{0.5cm}
Some care will need to be taken to normalize cross sections since one is
comparing Drell-Yan processes on protons and deuterons.  This should be
feasible by bombarding both hydrogen and deuterium targets simultaneously
with charged pion beams. Eq.\ (\ref{eq:Rnfinal}) assumes that deuteron
structure functions are just the sum of the free nucleon terms; if we include
corrections in the form of Eq.\ (\ref{eq:epsilon}), we obtain an additional
first-order correction
\begin{equation}
\delta R^{DY}_{\pi N}(x_1)  =  -\epsilon (x_1) \left( \frac {u^p + d^p }{d^p}
 \right) (x_1)
\label{eq:deltaR}
\end{equation}
This preserves the separation into nucleonic and pionic CSV terms, but depends
on `EMC' changes in the deuteron structure functions relative to free proton
and neutron distributions, and on the $u/d$ ratio of proton distributions.
For large $x$, $u(x)/d(x) >> 1$, so the EMC term could be significant even for
small values of $\epsilon (x)$. For $x \sim 0.5$, where $u^p/d^p \approx 4$,
if $\epsilon (x)$ is as large as -0.02 then $\delta R^{DY}_{\pi N}(x=0.5)
\approx 0.10$. At larger $x$ the EMC contribution could be even bigger, and
might conceivably dominate the CSV terms. Since all terms (pion and nucleon
CSV, and EMC) are predicted to have the same sign in the region $0.3 < x < 0.8$
(we expect $\epsilon (x) < 0$ in this region), the ratio $R^{DY}_{\pi N}$ could
be as large as 0.3.  In view of this sensitivity to the EMC term, it is
important that accurate calculations be carried out of Fermi motion and binding
corrections for the deuteron, including possible flavor dependence of such
corrections. Melnitchouk, Schreiber and Thomas \cite{mst,melnit} have recently
studied the contributions to $\epsilon (x)$ in the deuteron.

We have also calculated the sea-valence contribution to the Drell-Yan ratio
$R^{DY}_{\pi N}$.  Relative to the deuteron measurement, we predict a CSV
contribution which increases by about a factor 5.  The sea-valence background
also increases by about the same factor. So our remarks about the sea-valence
background and the CSV ``signal'' are equally valid for these Drell-Yan
processes.

In conclusion, we have shown that by comparing the Drell-Yan yield for $\pi^+$
and $\pi^-$ on nucleons or deuterons, one might be able to extract the charge
symmetry violating [CSV] parts of {\bf both} pion and nucleon.  We discussed
two different linear combinations of $\pi^+$ and $\pi^-$ induced Drell-Yan
cross sections, which produce a result directly proportional to the CSV terms.
Furthermore, we found that the ratio of Drell-Yan cross sections {\bf
separates}
completely into two terms, one of which ($R^N(x_N)$) depends only on the
nucleon CSV, and the other ($R^\pi(x_\pi)$) depends on the pion CSV
contribution. Thus if this ratio can be accurately measured as a function of
$x_N$ and $x_\pi$, both the nucleon and pion CSV terms might be extracted.
The largest background should arise from terms involving one sea quark and one
valence quark.  Such contributions, although relatively large, should be
predicted quite accurately, and may be subtracted off through their very
different behavior as a function of nucleon and pion $x$.  As the $x_1$ and
$x_2$ values of interest for the proposed measurements are large ($x > 0.5$),
a beam of 40-50 GeV pions will produce sufficiently massive dilepton pairs
that the Drell-Yan mechanism is applicable. A flux of more than $10^9$
pions/sec.\ is desirable, which may mean that the experiment is not feasible
until the new FNAL Main Ring Injector becomes operable.

%%%%%%%%%%%%%%%%%%%%%%%%%%%%%%%%%%%%%%%%%%%%%%%%%%%%%%%%%%%%%%%%%%%%%%%%%%%

This work was supported by the Australian Research Council.
One of the authors [JTL] was supported in part by the US NSF under research
contract NSF-PHY91-08036.


\begin{thebibliography}{99}

\bibitem{mrs} A.D.Martin, W.J.Stirling and R.G.Roberts,
		          {\em Phys. Lett.}{\bf B252} (1990) 653;\\
			  S.D.Ellis and W.J.Stirling,
			  {\em Phys. Lett. }{\bf B256} (1991) 258.

\bibitem{sst} A.Signal, A.W.Schreiber and A.W.Thomas,
             {\em Mod. Phys. Lett. }{\bf A6} (1991) 271.

\bibitem{henley} E.M.Henley and G.A.Miller,
			 {\em Phys. Lett. }{\bf B251} (1990) 453.

\bibitem{soffer} G.Preparata, P.G.Ratcliffe and J.Soffer,
			 {\em Phys. Rev. Lett. }{\bf 66} (1991) 687.

\bibitem{kum} S.Kumano,
		   {\em Phys. Rev. }{\bf D43} (1991) 59;\\
		   S.Kumano and J.T.Londergan,
		   {\em Phys. Rev. }{\bf D44} (1991) 717.

\bibitem{mt1} W.Melnitchouk and A.W.Thomas,
		   {\em Phys. Rev. }{\bf D47} (1993) 3783;\\
			V.R.Zoller,
			{\em Phys. Lett. }{\bf B279} (1992) 145;\\
			B.Badelek and J.Kwiecinski,
			{\em Nucl. Phys. }{\bf B370} (1991) 278.

\bibitem{mt2} W.Melnitchouk, A.W.Thomas and A.I.Signal,
			{\em Zeit. Phys. }{\bf 340} (1991) 85.

\bibitem{vary} M. Sawicki and J.P.Vary,
			{\em Phys. Rev. Lett. }{\bf 71} (1993) 1320.


\bibitem{nmc} P.Amaudruz et al. (NMC Collaboration),
			{\em Phys. Rev. Lett. }{\bf 66} (1991) 560.

\bibitem{emc} J. Ashman et al. (EMC Collaboration),
			{\em Phys. Lett.}{\bf B206} (1988) 364;\\
            {\em Nucl. Phys.}{\bf B328} (1990) 1.

\bibitem{slacspin} B. Adeva et al. (SMC Collaboration),
			{\em Phys. Lett.}{\bf B302} (1993) 533;\\
				   P. L. Anthony et al.,
            {\em Phys. Rev. Lett.}{\bf 71} (1993) 959.

\bibitem{revspin} R. Windmolders,
			{\em Int. J. Mod. Phys.}{\bf A7} (1992) 639;
				  S. D. Bass and A.W.Thomas,\\
            {\em J. Phys.}{\bf G19} (1993) 925.

\bibitem{sather} E. Sather,
			{\em Phys. Lett.}{\bf B274} (1992) 433.

\bibitem{rtl} E. Rodionov, A. W. Thomas and J. T. Londergan,
	     {\em Mod. Phys. Lett.}{\bf A9, 19} (1994) 1799.

\bibitem{mill} G. A. Miller, B. M. K. Nefkens and I. \v{S}laus,
			 {\em Phys. Rep. }{\bf 194} (1990) 1.

\bibitem{fec} F.E.Close,
		"An Introduction to Quarks and Partons"
		(Academic, London, 1979);\\
		E.Leader and E.Predazzi,
		"Gauge theories and the New Physics"
	    (Cambridge University Press, Cambridge, 1982).

\bibitem{mt3} A. W. Thomas and W. Melnitchouk,
		University of Adelaide preprint: ADP-93-217/T135,
		to appear in Proc. JSPS-INS Spring School, World Scientific
			(1994).

\bibitem{garvey} D. M. Alde et al.,
			 {\em Phys. Rev. Lett.}{\bf 64} (1990) 2479.

\bibitem{bodek} A. Bodek and J. L. Ritchie,
			 {\em Phys. Rev.}{\bf D23} (1981) 1070.

\bibitem{fs} L. L. Frankfurt and M. I. Strikman,
			 {\em Phys. Rep.}{\bf 160} (1988) 235.

\bibitem{bicker} R. P. Bickerstaff and A. W. Thomas,
			 {\em J. Phys.}{\bf G15} (1989) 1523.

\bibitem{mst} W. Melnitchouk, A. W. Schreiber and A. W. Thomas,
			 {\em Phys. Rev.}{\bf D49} (1994) 1183.

\bibitem{MIT} A. Chodos et al.
			 {\em Phys. Rev.}{\bf D10} (1974) 2599.

\bibitem{adel} A.I.Signal and A.W.Thomas,
			  {\em Phys.Rev. }{\bf D40} (1989) 2832;\\
               A.W.Schreiber, A.W.Thomas and J.T.Londergan,
			   {\em Phys. Rev. }{\bf D42} (1990) 2226.

\bibitem{birse} E. Naar and C. Birse,
				{\em Phys. Lett.}{\bf B305} (1993) 190.

\bibitem{altpar} G. Altarelli and G. Parisi,
			 {\em Nucl. Phys.}{\bf B126} (1977) 298.

\bibitem{close}    F.E.Close and A.W.Thomas,
				{\em Phys. Lett. }{\bf B212} (1988) 227.

\bibitem{far} G. Farrar and D. Jackson,
			 {\em Phys. Rev. Lett.} {\bf 35} (1975) 1416.

\bibitem{njl} Y. Nambu and G. Jona-Lasinio,
			 {\em Phys. Rev.}{\bf 122} (1961) 345;
			 {\it ibid.} {\bf 124} (1961) 246.

\bibitem{toki1} T. Shigetani, K. Suzuki and H. Toki,
			 {\em Phys. Lett.}{\bf B308} (1993) 383.

\bibitem{toki2} T. Shigetani, K. Suzuki and H. Toki,
			 Tokyo Metropolitan University preprint: TMU-NT940101
			 (1994).

\bibitem{approx} P.N. Harriman, A.D. Martin, W.J. Stirling and
		R.G. Roberts, Phys.\ Rev.\ {\bf D42} (1990) 798;
		P.J. Sutton, R.G. Roberts, A.D. Martin and W.J. Stirling,
		Phys.\ Rev.\ {\bf D45} (1992) 2349.  We used the nucleon
		parton distribution HMRS(B), and pion fit 3, for which
		the pion sea carries 10\% of the pion's momentum at
		$Q^2 = 4$ GeV$^2$.

\bibitem{melnit} W. Melnitchouk, A. W. Schreiber and A. W. Thomas,
			{\em Phys. Lett.}{\bf B 335} (1994) 11.

\end{thebibliography}
\end{document}